\documentclass[aps,pra,reprint,superscriptaddress,nofootinbib,longbibliography]{revtex4-2}
\usepackage{amsmath,amssymb,bm}
\usepackage{graphicx}
\usepackage{xcolor}
\usepackage{orcidlink}
\usepackage{hyperref}
\hypersetup{colorlinks=true,linkcolor=blue,citecolor=blue,urlcolor=blue}
\newcommand{\ket}[1]{\lvert #1\rangle}

\newcommand{\ketbra}[2]{\lvert #1\rangle\langle #2\rvert}
\newcommand{\ip}[2]{\langle #1\vert #2\rangle}
\newcommand{\Psc}{P_{\mathrm{sc}}}
\newcommand{\Oh}{\Omega_{\mathrm{hold}}}
\newcommand{\Ostray}{\Omega_{\mathrm{stray}}}
\newcommand{\Tr}{\mathrm{Tr}}
\newcommand{\Hnh}{H_{\mathrm{nh}}}

\begin{document}

\title{Reservoir-independent lossless charging and protected storage of an open quantum battery}

\author{Asad Ali\orcidlink{0000-0001-9243-417X}}
\email{asal68826@hbku.edu.qa}
\affiliation{Qatar Center for Quantum Computing, College of Science and Engineering, Hamad Bin Khalifa University, Doha, Qatar}

\author{H. Kuniyil\orcidlink{0000-0003-0338-1278}}
\affiliation{Qatar Center for Quantum Computing, College of Science and Engineering, Hamad Bin Khalifa University, Doha, Qatar}

\author{M.I Hussain\orcidlink{0000-0002-6231-7746}}
\affiliation{Qatar Center for Quantum Computing, College of Science and Engineering, Hamad Bin Khalifa University, Doha, Qatar}
\author{M.T Rahim\orcidlink{0000-0003-1529-928X}}
\affiliation{Qatar Center for Quantum Computing, College of Science and Engineering, Hamad Bin Khalifa University, Doha, Qatar}

\author{Saif Al-Kuwari\orcidlink{0000-0002-4402-7710}}
\affiliation{Qatar Center for Quantum Computing, College of Science and Engineering, Hamad Bin Khalifa University, Doha, Qatar}

\author{James~Q.~Quach\orcidlink{0000-0002-3619-2505}} 
\affiliation{The University of Adelaide, SA 5005, Australia}
\date{\today}

\begin{abstract}
A quantum battery charged through a lossy intermediate state faces a structural trade-off between charging speed and dissipation. We show that an exact algebraic cancellation removes it in a driven three-level cell: the radiatively decaying state is fed by a single bright amplitude, and a counterdiabatic field annuls the lone residual source that drives it, holding the lossy state identically empty. Charging is then lossless---not one photon is emitted through the bridge---at any one-photon detuning, coupling, linewidth, and speed down to the rotating-wave limit, with no adiabatic elimination, so the charging power is bounded by the drive amplitude (a quantum speed limit) rather than by dissipation. Crucially, this losslessness is independent of the reservoir: because the dark sector never engages the system--bath coupling, the emission vanishes exactly for an arbitrary spectral density, Markovian or not, as an exact damped-pseudomode treatment confirms to machine precision across all memory times. The entire non-Hermitian structure---a Markovian second-order exceptional point that reservoir memory promotes to a third-order one, and the attendant dissipation phase diagram---lives in the bright sector, from which the protocol is by construction exempt. This inverts dissipation-engineered charging, where an exceptional point or reservoir memory is a resource; here the lossy sector is never populated at all. The same dark-state structure protects the stored charge, converting fast radiative self-discharge into the slow metastable lifetime, with residuals quadratic in the control error. We detail experimental requirements and representative parameters for neutral alkaline-earth atoms, trapped ions, transmons, and defect centers.
\end{abstract}
\maketitle

\section{Introduction}
\label{sec:intro}

A quantum battery is a quantum system used to store energy and release it as
work~\cite{Alicki2013,CampaioliReview,RMPbatteries}. Its performance is fixed by
four quantities: the energy it stores, the fraction extractable as ordered
work---the \emph{ergotropy}~\cite{Allahverdyan2004}---the rate at which it
charges, and how well it holds charge against its environment. Much theoretical
effort has targeted charging power, where collective and globally entangling
operations charge an array faster than the same number of independent
cells~\cite{Binder2015,Campaioli2017,Ferraro2018,Andolina2019}. Because any real
cell is an open system, an equally central question is how dissipation limits
charging and storage~\cite{Farina2019,Barra2019,Carrega2020,Mitchison2021}, and a
large recent literature studies batteries in structured and non-Markovian
reservoirs, where reservoir memory can return energy to the system and thereby
aid performance~\cite{LiShenYi2022,SantosNM2020,XuEnv2021,Morrone2023,XuSelf2024,BarraPhase2022}.

The obstruction that motivates this work is structural. To charge quickly one
must drive strongly; strong driving of a two-level cell, or routing the transfer
through an auxiliary level, generically populates a short-lived excited state
that decays radiatively; and the energy meant for storage is then lost to the
electromagnetic field, the ergotropy degraded, and the cell heated. The sharpest
form of the tension is a question one can ask of an exactly solvable model:
\emph{must fast charging dissipate, or can a cell be charged quickly and
losslessly at once, and if so, how does the answer depend on the reservoir?}

We answer this with an exact result and draw from it a categorical distinction
from existing approaches. In a driven three-level $\Lambda$ cell, the
radiatively decaying state is sourced by a \emph{single} ``bright'' combination
of the two storage amplitudes; we call this the \emph{source identity}. A single
counterdiabatic control field cancels the lone residual coupling---the
\emph{residual source}---that feeds the bright amplitude, after which the lossy
state stays \emph{identically} empty. Charging is then lossless: not a single
photon is emitted, the cell reaches full ergotropy, and the charging power is
bounded by the available drive amplitude---a quantum speed
limit~\cite{Mandelstam1945,Margolus1998,Deffner2017}---rather than by the
linewidth, at any one-photon detuning, coupling, and speed down to the rotating-wave limit (with the two-photon detuning held at $\delta=0$),
with the excited state retained in full and no adiabatic elimination.

The central new statement of this paper is that this losslessness is
\emph{independent of the reservoir}. Because the protocol holds the system in a
dark state with no excited-state component, the system--bath coupling operator
annihilates the state at every instant: the reservoir is never excited, so the
emission vanishes exactly for an \emph{arbitrary} bath spectral density, whether
Markovian or strongly non-Markovian, and whether or not that bath would, on its
own, place the lossy sector at an exceptional point. We confirm this with an
exact damped-pseudomode treatment of a Lorentzian reservoir, finding
machine-precision losslessness across the entire range of memory times, and we
show that all of the non-Hermitian structure---a Markovian second-order
exceptional point that reservoir memory promotes to a third-order one, and the
attendant dissipation phase diagram---lives in the bright sector, from which the
lossless protocol is by construction exempt.

This is the opposite philosophy to dissipation-engineered charging. Recent
proposals deliberately engineer a reservoir so that an exceptional point or the
broken phase amplifies the charging dynamics~\cite{ReservoirEP2025,Nonreciprocal2025},
or exploit reservoir memory and backflow to enhance ergotropy, power, or
self-discharge~\cite{LiShenYi2022,SantosNM2020,XuEnv2021,XuSelf2024,BarraPhase2022};
in those schemes dissipation is a \emph{resource}. Here it is an
\emph{irrelevance}: the protocol is defined by never populating the lossy sector,
so memory and exceptional points have nothing to act on. The exceptional-point
phenomenology that has recently been observed in non-Markovian open
systems~\cite{NMEP2025,OptoNMEP2026} is present in our cell---and we map it---but
the charging protocol sits on the sheet where it produces no dissipation at all.

The dark-state idea itself is not new to batteries: dark states have been used to
charge and stabilize open cells~\cite{Quach2020,Liu2019,Tejero2021} and to build
stable adiabatic batteries~\cite{Santos2019,Santos2020}, and the optical route
rests on stimulated Raman adiabatic passage
(STIRAP)~\cite{Gaubatz1990,Vitanov2017,Bergmann2019}, accelerated by shortcuts to
adiabaticity in closed systems~\cite{Demirplak2003,Berry2009,Chen2010,Giannelli2014,Vepsalainen2019,GueryOdelin2019}.
What is new here is that the bright-source decomposition makes the open-system
charging \emph{exactly} lossless at \emph{any} speed and, crucially,
\emph{independent of the reservoir}, whereas those approaches are adiabatic
(hence slow), rate-limited by the environment, or constructed for unitary
dynamics and analyzed by eliminating the lossy state.

The bright-source/residual-source decomposition that makes this exact and
transparent also supplies the remaining results: the suppression of dissipation
under an imperfect control is quadratic in the calibration error; the same dark
direction, actively dressed by a holding field, protects the stored charge,
converting a fast radiative self-discharge into the slow intrinsic metastable
lifetime; and when the storage transition is dipole-forbidden and the control
must be synthesized as a two-photon Raman process, the exact $\Psc=0$ is replaced
by a small, quantified residual. We close with the experimental requirements and
representative parameters for neutral alkaline-earth atoms, trapped ions,
superconducting transmons, and defect centers.

The paper is organized as follows. Section~\ref{sec:model} defines the cell and
the figures of merit. Section~\ref{sec:bsc} establishes the bright-source
decomposition, and Sec.~\ref{sec:lossless} proves the exact lossless-charging
theorem. Section~\ref{sec:power} bounds the power by the drive, and
Sec.~\ref{sec:robust} treats robustness. Section~\ref{sec:reservoir} is the
central one: the reservoir independence of the losslessness, its exact
non-Markovian verification, the exceptional-point phase diagram of the bright
sector, and the categorical distinction from dissipation-engineered batteries.
Section~\ref{sec:storage} develops protected storage, Sec.~\ref{sec:raman} the
Raman-synthesis cost, and Sec.~\ref{sec:platforms} the experimental realization.
Section~\ref{sec:discussion} discusses scope and Sec.~\ref{sec:conclusion}
concludes. Appendices give the no-jump derivation, the pseudomode construction,
the exceptional-point spectral analysis, and the protected-storage algebra.

\begin{figure}[t]
\centering
\includegraphics[width=0.9\columnwidth]{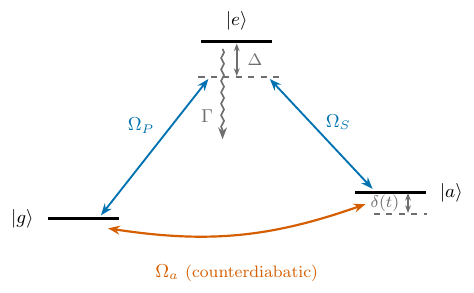}
\caption{\label{fig:cell}\textbf{The three-level battery cell.} Long-lived states
$\ket g$ (empty) and $\ket a$ (charged, stored energy $\hbar\omega_a$) are coupled
to a radiatively decaying bridge $\ket e$ (linewidth $\Gamma$) by a pump
$\Omega_P$ on $\ket g$--$\ket e$ and a Stokes field $\Omega_S$ on $\ket a$--$\ket e$
at one-photon detuning $\Delta$ (dashed: the virtual level where the optical
fields terminate). A counterdiabatic control $\Omega_a$ (red) drives the
metastable $\ket g$--$\ket a$ transition directly. Charging transfers population
$\ket g\to\ket a$; the control keeps $\ket e$ exactly empty, so the charge is
delivered with no photon emitted.}
\end{figure}

\section{Battery cell and figures of merit}
\label{sec:model}

\subsection{Cell, stored energy, and ergotropy}

The cell is a three-level $\Lambda$ system (Fig.~\ref{fig:cell}) with two
long-lived states $\ket g$ and $\ket a$ and a radiatively decaying state $\ket e$
of linewidth $\Gamma$. Taking $\ket g$ at zero energy, the battery Hamiltonian is
\begin{equation}
H_B=\hbar\omega_a\,\ketbra a a+\hbar\omega_e\,\ketbra e e ,
\qquad \omega_e\gg\omega_a .
\label{eq:HB}
\end{equation}
The charged state is $\ket a$: the stored energy is the population transferred
from $\ket g$ to $\ket a$, of energy $\hbar\omega_a$, while $\ket e$ serves only
as an optical bridge. Physically $\ket a$ is a metastable or optical-clock state
with a long intrinsic lifetime, and $\ket e$ is an ordinary optically excited
state.

The operationally relevant figure of merit is the ergotropy, the maximum work
extractable by a cyclic unitary~\cite{Allahverdyan2004},
\begin{equation}
\mathcal W(\rho)=\max_{U}\big[\Tr(\rho H_B)-\Tr(U\rho U^\dagger H_B)\big].
\label{eq:ergotropy}
\end{equation}
The optimal $U$ reorders populations to be anti-aligned with the energies,
producing the passive state; for a pure state the passive state is the ground
state~\cite{Pusz1978,Lenard1978}, so all of the energy of a pure excited state is
extractable. A mixed state---in particular one produced by partial decay during
charging---has ergotropy strictly below its stored energy, because part of the
energy is locked in disorder. Dissipation during charging is thus doubly
damaging: it removes energy and degrades what remains. The protocol below
delivers the cell in the pure charged state $\ket a$, so its ergotropy is
maximal, $\mathcal W=\hbar\omega_a$.

\subsection{Drives, dissipation, and the emitted-photon probability}

A pump field of Rabi frequency $\Omega_P$ drives $\ket g$--$\ket e$, a Stokes
field $\Omega_S$ drives $\ket a$--$\ket e$, and a control field $\Omega_a$ drives
the metastable $\ket g$--$\ket a$ transition directly; $\Delta$ and $\delta$ are
the one- and two-photon detunings. Writing
$\ket{\psi}=c_g\ket g+c_a\ket a+c_e\ket e$ in the rotating frame, the Hamiltonian
within the rotating-wave approximation
($\Omega_P,\Omega_S,\Omega_a\ll\omega_a,\omega_e$) is, with $\hbar=1$,
\begin{equation}
\begin{split}
H={}&\frac{\Omega_P}{2}\ketbra e g+\frac{\Omega_S}{2}\ketbra e a
+\frac{i\Omega_a}{2}\big(\ketbra g a-\ketbra a g\big)\\
&-\Delta\,\ketbra e e-\delta\,\ketbra a a+\mathrm{h.c.}
\end{split}
\label{eq:H}
\end{equation}
Spontaneous emission into the vacuum is described by the Lindblad master equation
$\dot\rho=-i[H,\rho]+\Gamma\big(\ketbra g e\,\rho\,\ketbra e g
-\tfrac12\{\ketbra e e,\rho\}\big)$. In the quantum-trajectory
unravelling~\cite{PlenioKnight1998,BreuerPetruccione}, the evolution between
emissions is generated by the non-Hermitian Hamiltonian
$\Hnh=H-\tfrac{i}{2}\Gamma\ketbra e e$, and the probability that a photon has been
emitted---the dissipated fraction of the charge---is the time-integrated
excited-state population,
\begin{equation}
\Psc=\Gamma\int_0^T|c_e(t)|^2\,dt ,
\label{eq:psc}
\end{equation}
with $c_e=\ip{e}{\psi}$ (Appendix~\ref{app:nojump}). Equation~\eqref{eq:psc} is
exact and makes the design goal sharp: controlling dissipation means controlling
$c_e$, and if $c_e(t)$ is held at zero during a fast charge, no photon is emitted
and the charge is delivered in full.

\section{Bright-source cancellation}
\label{sec:bsc}

\subsection{Dark/bright basis decomposition and the source identity}

With the mixing angle $\tan\theta=\Omega_P/\Omega_S$ and
$\Omega=\sqrt{\Omega_P^2+\Omega_S^2}$, define the dark and bright states (see Appendix~\ref{app:darkbasis} to distinguish it from dark and bright basis)
\begin{equation}
\ket D=\cos\theta\,\ket g-\sin\theta\,\ket a,\qquad
\ket B=\sin\theta\,\ket g+\cos\theta\,\ket a .
\label{eq:db}
\end{equation}
The optical couplings in Eq.~\eqref{eq:H} annihilate $\ket D$ while $\ket B$
retains a coupling of strength $\Omega/2$ to $\ket e$: an amplitude in $\ket D$
never reaches the lossy state; an amplitude in $\ket B$ has a doorway to it. As
$\theta$ sweeps from $0$ to $\pi/2$ (Stokes before pump), $\ket D$ rotates from
$\ket g$ to $-\ket a$, so following the dark state carries the cell from empty to
charged through a state that never occupies $\ket e$. In terms of the dark and
bright amplitudes $d=\ip{D}{\psi}$ and $b=\ip{B}{\psi}$, this is made exact by the
\emph{source identity}
\begin{equation}
\Omega_P\,c_g+\Omega_S\,c_a=\Omega\,b ,
\label{eq:source}
\end{equation}
which states that the only combination of the $\ket g,\ket a$ amplitudes that
feeds the lossy state is the bright amplitude.

\subsection{Equations of motion and the residual source}

At two-photon resonance ($\delta=0$) the dynamics closes on $(d,b,c_e)$ as
\begin{equation}
i\dot d=-iF\,b,\qquad
i\dot b=iF\,d+\frac{\Omega}{2}c_e,\qquad
i\dot c_e=\frac{\Omega}{2}b-\zeta\,c_e ,
\label{eq:eom}
\end{equation}
with $\zeta=\Delta+\tfrac{i}{2}\Gamma$ and the \emph{residual source}
\begin{equation}
F=\dot\theta-\frac{\Omega_a}{2}.
\label{eq:F}
\end{equation}
Equation~\eqref{eq:eom} is the heart of the construction. The lossy amplitude
$c_e$ is driven \emph{only} by $b$; the dark amplitude $d$ never couples to
$\ket e$ directly; and the protected dark sector and the lossy bright--excited
sector are joined by the \emph{single} quantity $F$. All dissipation flows
through $\ket B$, while $\ket D$ is intrinsically loss-free. For nonzero $\delta$,
the residual source acquires an imaginary part,
$\boldsymbol{F}=F-i\,\delta\sin\theta\cos\theta$, whose imaginary
quadrature is proportional to $\delta$ and \emph{independent} of $\Omega_a$;
exact cancellation therefore requires $\delta=0$. The control can compensate the
non-adiabatic (real) mismatch but is structurally blind to a two-photon or
Doppler detuning, which lives in the orthogonal quadrature ($\operatorname{Im} \boldsymbol{F}$).

\section{Exact lossless charging}
\label{sec:lossless}

\subsection{Counterdiabatic field and the cancellation theorem}

Adiabatic following keeps the cell in $\ket D$ and never excites $\ket e$, but
demands a slow sweep: with $\Omega_a=0$ the residual source is $F=\dot\theta$,
and a fast sweep makes it large, driving $b$ and hence $c_e$, which radiates. The
remedy is to cancel the residual source. Choosing the \emph{counterdiabatic
field}
\begin{equation}
\Omega_a(t)=2\dot\theta(t)\qquad\Longrightarrow\qquad F\equiv0
\label{eq:cd}
\end{equation}
removes the dark--bright coupling at every instant; this is the
transitionless-driving prescription of counterdiabatic
STIRAP~\cite{Demirplak2003,Berry2009,Chen2010,GueryOdelin2019}, read off the
open-system equations. With $F\equiv0$ the dark sector decouples exactly.
Starting from $\ket g=\ket{D(0)}$ one has $b(0)=c_e(0)=0$, and the
bright--excited pair $i\dot b=\tfrac{\Omega}{2}c_e$,
$i\dot c_e=\tfrac{\Omega}{2}b-\zeta c_e$ is linear and homogeneous with zero
initial data; by uniqueness of solutions, $b(t)\equiv0$ and $c_e(t)\equiv0$ for
all $t$, independently of $\Delta$, $\Omega$, $\Gamma$, and the charging time $T$.
Consequently, by Eq.~\eqref{eq:psc},
\begin{equation}
\Psc=\Gamma\int_0^T|c_e|^2\,dt=0\quad\text{exactly}.
\label{eq:theorem}
\end{equation}
Not a single photon is emitted \emph{through the bridge} $\ket e$: the cell
follows $\ket{D(t)}$ from $\ket g$ to $\ket a$ and ends in the pure state $\ket a$
of energy $\hbar\omega_a$. (Equation~\eqref{eq:psc} counts emission through the
radiative bridge, the only fast channel; the metastable state $\ket a$ decays at
its slow intrinsic rate $\gamma_a$, which is a separate channel and is negligible
over the charging window $T\ll\gamma_a^{-1}$.) 

\begin{figure*}[t]
\centering
\includegraphics[width=\textwidth]{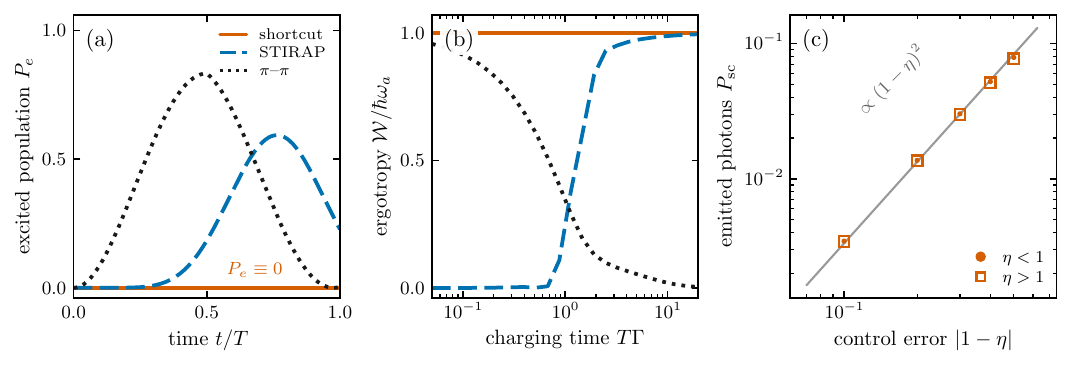}
\caption{\label{fig:charge}\textbf{Lossless charging.} Population is transferred
$g\!\to\!a$ (stored energy $\hbar\omega_a$) through the lossy state $\ket e$ under
the full master equation; the cell is relaxed before the extractable work is read
out. Counterdiabatic shortcut $\Omega_a=2\dot\theta$ (orange), bare STIRAP
(blue), and the intuitive $\pi$--$\pi$ sequence $g\!\to\!e\!\to\!a$ (black).
(a)~Excited population at charging time $T\Gamma=1$: the shortcut keeps the bright
source cancelled, so $\ket e$ stays exactly empty, $P_e\equiv0$. (b)~Settled
ergotropy versus charging time: full extractable work at every speed for the
shortcut, slow-only for STIRAP, fast-only for the $\pi$--$\pi$ sequence. (c)~With
an imperfect shortcut $\Omega_a=2\eta\dot\theta$ the residual source is
$F=(1-\eta)\dot\theta$, and the dissipated energy collapses onto a single
$(1-\eta)^2$ law for $\eta\lessgtr1$ (slope-2 line). Parameters $\Omega=10\Gamma$,
$\Delta=0$, $\omega_e=5\omega_a$.}
\end{figure*}

The statement is exact, with $\ket e$ retained in full---no adiabatic elimination and
no expansion in any small parameter---and it holds at any speed down to the
rotating-wave limit. It is the open-system counterpart of counterdiabatic
STIRAP: there the added $\ket g$--$\ket a$ field cancels unitary non-adiabatic
transitions; here the \emph{same} field annuls the residual source, and thereby
the emission itself.

\subsection{Ergotropy and numerical verification}

Because the protocol delivers the pure charged state $\ket a$, its ergotropy
[Eq.~\eqref{eq:ergotropy}] equals the full stored energy,
$\mathcal W=\hbar\omega_a$, the maximum possible. Figure~\ref{fig:charge} verifies
the theorem by integrating the full master equation: the counterdiabatic protocol
reaches $\mathcal W=\hbar\omega_a$ to numerical precision with fewer than
$10^{-11}$ emitted photons, at every charging time, while the excited population
$P_e=|c_e|^2$ stays flat at zero.

\section{Charging power limited by drive, not loss}
\label{sec:power}

The charging power is $P=\hbar\omega_a/T$. Because the shortcut is lossless (for
an independent $\ket g$--$\ket a$ drive), $P$ is bounded not by dissipation but by
the control. The counterdiabatic amplitude needed to sweep $\theta$ through
$\pi/2$ in time $T$ scales as $\Omega_a=2\dot\theta\sim\pi/T$, so the achievable
power is set by the available Rabi frequency,
\begin{equation}
P\sim\frac{\hbar\omega_a\,\Omega_a}{\pi}.
\label{eq:power}
\end{equation}
This is a quantum speed limit: the minimum time to drive $\ket g$ to the
orthogonal state $\ket a$ is fixed by the energy scale of the generator that
connects them~\cite{Mandelstam1945,Margolus1998,Deffner2017}, and the dark-state
shortcut saturates that limit without dissipating. Here the relevant generator is
the controlled $\ket g$--$\ket a$ coupling of amplitude $\Omega_a$, itself bounded
by the rotating-wave condition $\Omega_a\ll\omega_a$; the bound~\eqref{eq:power} is
thus the speed limit set by the \emph{available control}, not the larger one set
by the stored-energy scale $\hbar\omega_a$ itself. The familiar trade-off in
which faster charging costs more dissipation is thereby revealed to be an
artifact of routing population through the lossy state. The treatment lives
within the rotating-wave approximation, $\Omega,\Omega_a\ll\omega_a,\omega_e$,
which bounds the charging time from below, $T\gtrsim\pi/\omega_a$.
Figure~\ref{fig:charge} compares three protocols at $\Omega=10\Gamma$: the
$\pi$--$\pi$ sequence charges fast but fills $\ket e$ and dissipates unless driven
faster than $\Gamma$; bare adiabatic passage avoids $\ket e$ only when slow; the
shortcut alone delivers full ergotropy at every speed, the unique optical
protocol that is simultaneously fast and lossless.

\section{Robustness to control errors}
\label{sec:robust}

\subsection{Amplitude miscalibration}

Perfect cancellation requires $\Omega_a=2\dot\theta$ exactly. For a miscalibrated
control $\Omega_a=2\eta\dot\theta$ ($\eta\neq1$), Eq.~\eqref{eq:F} gives a
residual source $F=(1-\eta)\dot\theta$ that no longer vanishes, so the dark state
leaks into the bright--excited sector, $b$ and hence $c_e$ acquire amplitudes of
order $(1-\eta)$, and by the quadratic form of Eq.~\eqref{eq:psc} the emitted
energy and the ergotropy deficit grow as
\begin{equation}
\Psc\propto(1-\eta)^2 ,
\label{eq:robust}
\end{equation}
the same law for under- and over-driving [Fig.~\ref{fig:charge}(c)]. A
one-percent calibration error costs only of order $10^{-4}$ in dissipated charge;
the protocol is forgiving, and the lossless limit returns smoothly as
$\eta\to1$.
\begin{figure*}[t]
\centering
\includegraphics[width=\textwidth]{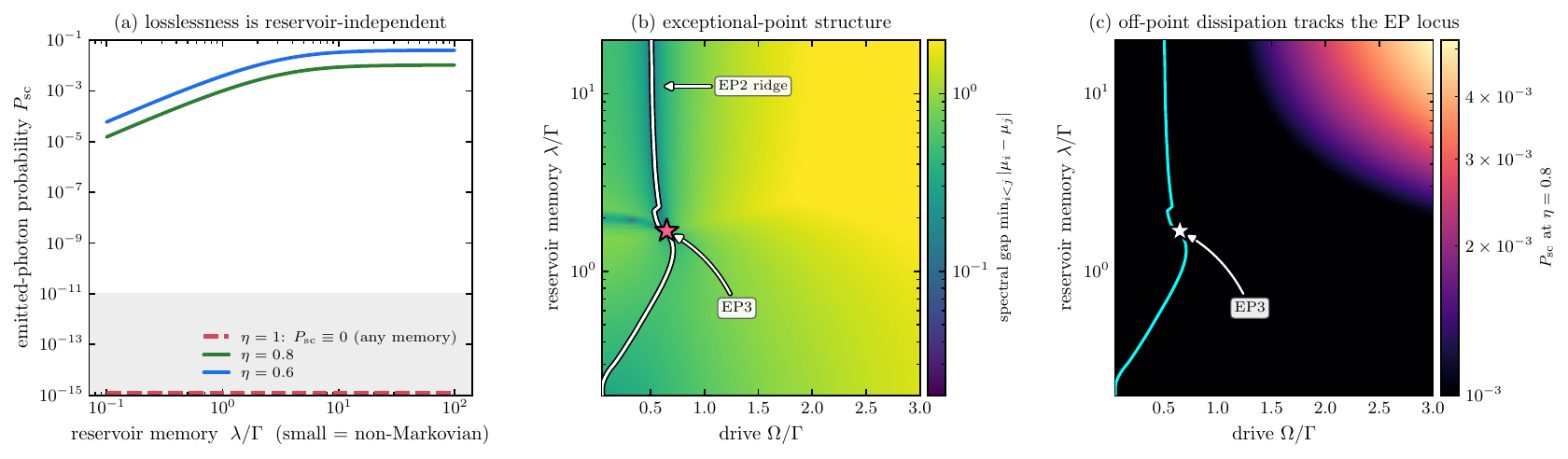}
\caption{\label{fig:reservoir}\textbf{Reservoir independence and the
exceptional-point structure of the bright sector.} A Lorentzian reservoir of
width $\lambda$ on $\ket e$ is treated exactly as a damped pseudomode
[Eq.~\eqref{eq:pseudo}]; $\Gamma=2g^2/\lambda$ is the Markovian rate.
(a)~Emitted-photon probability versus memory $\lambda/\Gamma$ (small $=$
non-Markovian) at $\Omega=10\Gamma$, $T\Gamma=1$. On the cancellation point
($\eta=1$, dashed red) $\Psc\equiv0$ to machine precision for every memory time;
off the point ($\eta=0.8,0.6$) the loss is finite and is \emph{suppressed} by
reservoir memory (backflow). (b)~Spectral gap $\min_{i<j}|\mu_i-\mu_j|$ of the
bright--excited--pseudomode generator over the $(\Omega/\Gamma,\lambda/\Gamma)$
plane. The Markovian second-order exceptional point at $\Omega=\Gamma/2$ (dotted)
is the large-$\lambda$ end of an exceptional ridge (white) that, at finite memory,
passes through a third-order exceptional point (star) at
$(\Omega/\Gamma,\lambda/\Gamma)=(3\sqrt3/8,\,27/16)\approx(0.65,1.69)$
(Appendix~\ref{app:ep}). (c)~Off-point dissipation $\Psc$ at $\eta=0.8$ over the
same plane, tracking the exceptional ridge (cyan); the on-point sheet ($\eta=1$)
is identically zero everywhere on this plane.}
\end{figure*}

 This quadratic insensitivity is what makes the scheme realistic, as
it tolerates the amplitude and timing imperfections of real pulse generators.

\subsection{Two-photon detuning}
\label{sec:detuning}

\begin{figure*}[t]
\centering
\includegraphics[width=0.95\textwidth]{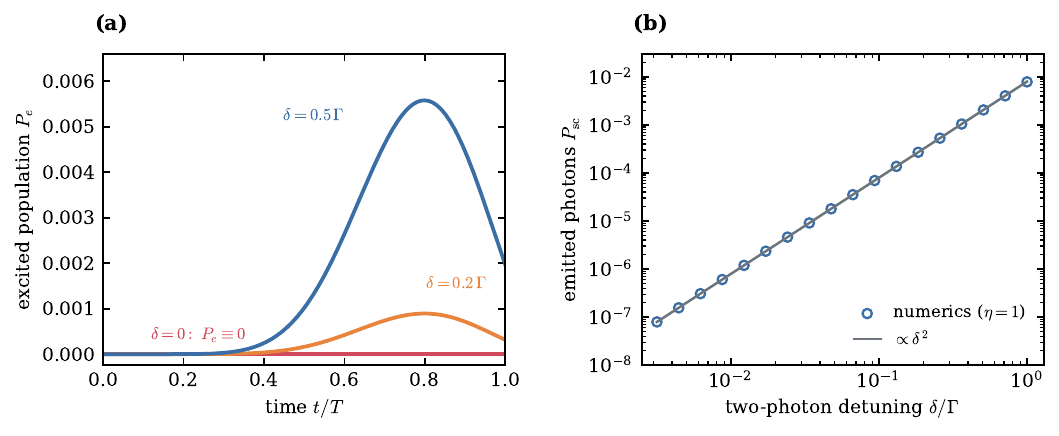}
\caption{\label{fig:detuning}\textbf{Sensitivity to two-photon detuning.} On the
amplitude-matched cancellation point ($\Omega_a=2\dot\theta$, i.e.\ $\eta=1$) a
finite two-photon detuning $\delta$ feeds the uncancellable imaginary quadrature
of the residual source $\operatorname{Im}F=-\delta\sin\theta\cos\theta$.
(a)~Excited population $P_e(t)$ at $T\Gamma=1$, $\Omega=10\Gamma$: at $\delta=0$ it
is identically zero, while $\delta=0.2\Gamma,0.5\Gamma$ raise a bright transient.
(b)~Emitted-photon probability $\Psc$ versus $\delta/\Gamma$ follows the
$\propto\delta^2$ law (line) over five decades, reaching machine zero at exact
two-photon resonance. Unlike the amplitude error of
Fig.~\ref{fig:charge}(c), this channel is structurally blind to the
counterdiabatic control.}
\end{figure*}

The amplitude error above lives in the \emph{real} quadrature of the residual
source and is therefore cancellable, in principle, by the control. The
qualitatively distinct imperfection is a finite two-photon (or Doppler) detuning
$\delta$, which enters the orthogonal quadrature: from Sec.~\ref{sec:bsc},
$F=(\dot\theta-\Omega_a/2)-i\,\delta\sin\theta\cos\theta$, so even on the
amplitude-matched point $\Omega_a=2\dot\theta$ the imaginary part
$\operatorname{Im}F=-\delta\sin\theta\cos\theta$ survives and is, by construction,
\emph{independent of $\Omega_a$}---no choice of the control field can remove it.
This residual source drives the bright amplitude, and hence $c_e$, with an
amplitude of order $\delta$, so by the quadratic form of Eq.~\eqref{eq:psc} the
emission grows as
\begin{equation}
\Psc\propto\delta^2 ,
\label{eq:detuning}
\end{equation}
the dominant practical imperfection for free atoms and ions, where residual
Doppler shifts and light shifts set the floor on $\delta$.
Figure~\ref{fig:detuning} confirms this: panel (a) shows the excited population
held identically at zero for $\delta=0$ while a finite $\delta$ raises a bright
transient, and panel (b) shows $\Psc$ following the $\delta^2$ law over five
decades down to machine zero at $\delta=0$. The scaling is benign---a detuning of
$10^{-2}\Gamma$ costs $\sim\!10^{-6}$ in dissipated charge---but, unlike the
amplitude error, it cannot be removed by the counterdiabatic field and so defines
the precision to which two-photon resonance must be held.

\section{Reservoir independence}
\label{sec:reservoir}

The results so far used the Markovian (flat-spectrum) form of the dissipation. We
now establish the central structural claim of the paper: the losslessness does
not depend on that assumption, or on any property of the reservoir at all.

\subsection{The losslessness holds for any reservoir}
\label{sec:bathagnostic}

Couple $\ket e$ to a general bosonic reservoir in its vacuum through the radiative
interaction
\begin{equation}
H_{SB}=\sum_k\big(g_k\,\ketbra g e\,b_k^\dagger+g_k^{*}\,\ketbra e g\,b_k\big),
\label{eq:HSB}
\end{equation}
with arbitrary couplings $\{g_k\}$, i.e.\ an arbitrary spectral density
$J(\omega)=\sum_k|g_k|^2\delta(\omega-\omega_k)$; the Markovian master equation of
Sec.~\ref{sec:model} is the flat-$J$ limit, but here $J$ is unrestricted. The
joint initial state is $\ket{\psi(0)}\otimes\ket{\mathrm{vac}}$ with
$\ket{\psi(0)}=\ket g$. On the cancellation point ($\Omega_a=2\dot\theta$,
$\delta=0$), the system amplitude follows the dark state $\ket{D(t)}$ exactly,
because the dark sector decouples (Sec.~\ref{sec:lossless}) and starts in
$\ket{D(0)}=\ket g$. Since $\ip{e}{D(t)}=0$ at every instant, the
interaction~\eqref{eq:HSB} acting on the joint state gives
\begin{equation}
H_{SB}\,\ket{D(t)}\otimes\ket{\mathrm{vac}}
=\sum_k g_k\,\ip{e}{D(t)}\,\ket g\otimes b_k^\dagger\ket{\mathrm{vac}}=0 .
\end{equation}
No reservoir excitation is ever created; the joint state remains
$\ket{D(t)}\otimes\ket{\mathrm{vac}}$ for all $t$; and the emitted-photon number,
equal to the number of reservoir excitations, is exactly zero,
\begin{equation}
\Psc=0\quad\text{for any }J(\omega),\ \text{Markovian or not}.
\label{eq:bathfree}
\end{equation}
This is stronger than Eq.~\eqref{eq:theorem}: it does not invoke the Markov
approximation or any unravelling, but follows directly from the joint
system--reservoir wavefunction. The losslessness is therefore immune to the
reservoir's memory, to its spectral structure, and to whether that structure
would place the lossy sector at an exceptional point. The dark state is an exact
dark state of the full system--reservoir coupling, not merely of the optical
drives. The assumptions are explicit and physical: the bath couples to the system
only through the $\ket e$ transition, the bath is in vacuum (zero temperature),
and the control is exact ($\Omega_a=2\dot\theta$, $\delta=0$) with the cell
started in $\ket g$.

\subsection{Exact non-Markovian verification: a damped pseudomode}
\label{sec:pseudomode}

To make Eq.~\eqref{eq:bathfree} concrete and to study the dynamics \emph{off} the
cancellation point, we replace the flat reservoir by a Lorentzian one of width
$\lambda$ centered on the $\ket e$ transition. Such a reservoir is reproduced
\emph{exactly} by coupling $\ket e$ to a single damped harmonic
pseudomode~\cite{Garraway1997} (Appendix~\ref{app:pseudomode}): in the
single-excitation, no-jump sector the bright amplitude $b$, the excited amplitude
$c_e$, and the pseudomode amplitude $A$ obey, at $\Delta=0$,
\begin{equation}
i\dot b=iF d+\frac{\Omega}{2}c_e,\quad
i\dot c_e=\frac{\Omega}{2}b+gA,\quad
i\dot A=g\,c_e-i\lambda A ,
\label{eq:pseudo}
\end{equation}
with system--pseudomode coupling $g$ and the Markovian effective rate recovered
as $\Gamma=2g^2/\lambda$ when $\lambda\to\infty$. The emitted-photon probability
is the norm lost from the pseudomode,
$\Psc=2\lambda\int_0^T|A|^2dt=1-\big(|d|^2+|b|^2+|c_e|^2+|A|^2\big)\big|_T$, and the
memory time is $\lambda^{-1}$: large $\lambda/\Gamma$ is Markovian, small
$\lambda/\Gamma$ is strongly non-Markovian with reservoir backflow.

On the cancellation point ($F=0$, started dark) the homogeneous
system~\eqref{eq:pseudo} has zero initial data, so $b\equiv c_e\equiv A\equiv0$
and $\Psc=0$ for \emph{any} $g$ and $\lambda$, confirming
Eq.~\eqref{eq:bathfree}. Figure~\ref{fig:reservoir}(a) shows $\Psc$ remaining at
machine zero across three decades of memory time, while away from the point
($\eta\neq1$) the dissipation is nonzero and \emph{decreases} as the memory
lengthens, the reservoir-backflow improvement familiar from non-Markovian
batteries~\cite{LiShenYi2022,XuEnv2021}. The lossless protocol neither needs nor
uses that improvement; it is already exact.

\subsection{Off the point: exceptional points and the phase diagram}
\label{sec:ep}

All of the non-Hermitian structure of the cell lives in the bright sector that
the protocol keeps empty. The homogeneous generator of $(b,c_e,A)$ in
Eq.~\eqref{eq:pseudo} is the non-Hermitian matrix
\begin{equation}
M_3=\begin{pmatrix}0 & \Omega/2 & 0\\ \Omega/2 & 0 & g\\ 0 & g & -i\lambda\end{pmatrix},
\label{eq:M3}
\end{equation}
whose three complex frequencies are the roots of
$\mu^3+i\lambda\mu^2-(g^2+\Omega^2/4)\mu-i\lambda\Omega^2/4=0$. In the Markovian
limit the pseudomode is eliminated and $M_3$ reduces to the damped two-level core
$M_2=\big(\begin{smallmatrix}0 & \Omega/2\\ \Omega/2 & -i\Gamma/2\end{smallmatrix}\big)$,
which has a second-order exceptional point (EP2) at $\Omega=\Gamma/2$, where the
two eigenvalues and eigenvectors coalesce, the generator becomes a Jordan block,
and a residual excited amplitude decays at the critical-damping rate
$c_e\propto t\,e^{-\Gamma t/4}$ (Appendix~\ref{app:ep}). A structured reservoir
adds a degree of freedom and enriches this: the EP2 becomes a ridge in the
$(\Omega/\Gamma,\lambda/\Gamma)$ plane that, at finite memory, passes through a
\emph{third-order} exceptional point (EP3) at $\Omega=2\lambda/3\sqrt3$,
$g^2=8\lambda^2/27$---equivalently $\lambda=27\Gamma/16$, $\Omega\approx0.65\Gamma$---where
all three eigenvalues coalesce and the critical transient becomes
$t^2\,e^{-\lambda t/3}$. This is precisely the promotion of a Markovian EP2 to a
non-Markovian EP3 by reservoir memory that has recently been observed in
circuit-QED~\cite{NMEP2025} and analyzed in optomechanics~\cite{OptoNMEP2026}.
Figure~\ref{fig:reservoir}(b) maps the exceptional ridge and the EP3, and
Fig.~\ref{fig:reservoir}(c) shows the off-point dissipation tracking this
structure across the plane. The decisive point is the contrast between panels:
the off-point dissipation has the full exceptional-point landscape, while the
on-point sheet is identically zero everywhere on the same plane. The cell
contains the exceptional-point physics; the lossless protocol simply does not
visit it.

\subsection{Categorical distinction from dissipation-engineered batteries}
\label{sec:distinction}

This clarifies how the present scheme differs in kind, not degree, from recent
open-system battery proposals. One line deliberately engineers a reservoir so
that an exceptional point or the resulting broken phase amplifies the charging
dynamics: reservoir-engineered exceptional points drive the stored energy to grow
exponentially under a bounded drive in the broken phase~\cite{ReservoirEP2025},
and a nonreciprocal battery operated at an exceptional point gains charging power
and resilience to parameter fluctuations~\cite{Nonreciprocal2025}. A second line
exploits reservoir memory and information backflow to enhance ergotropy, charging
power, or self-discharge in non-Markovian
environments~\cite{LiShenYi2022,SantosNM2020,XuEnv2021,XuSelf2024,BarraPhase2022}.
In both lines dissipation, or its memory, is a \emph{resource} to be harnessed,
and the battery operates \emph{within} the lossy, non-Hermitian sector where that
resource lives.

The protocol here is the categorical opposite. It is defined by never populating
the lossy sector at all: the cell rides a dark state that is decoupled from the
bright--excited--reservoir manifold, so the exceptional points, the broken phase,
and the reservoir memory have nothing to act on. The losslessness is consequently
not approximate, not bath-tuned, and not speed-limited by relaxation; it is exact
and reservoir-independent [Eq.~\eqref{eq:bathfree}]. Where dissipation-engineered
schemes ask the environment to help charge the battery, this one charges so that
the environment is never engaged. The two strategies are complementary---one
could imagine handing a charged cell from this protocol to an exceptional-point
extraction stage---but they occupy opposite regimes of the same open-system
phase diagram, and only the dark-state route makes the charging itself provably
lossless for an arbitrary reservoir.

\section{Protected storage}
\label{sec:storage}

A charged cell must hold its energy. Once stored, $\ket a$ is metastable and
decays only at its slow intrinsic rate $\gamma_a$, but any residual coupling back
to $\ket e$---imperfect extinction of the Stokes beam, stray resonant
light---reopens the fast radiative channel and discharges the cell on the scale
of $\Gamma^{-1}\ll\gamma_a^{-1}$. The bright-source structure again protects.
Suppose a parasitic field of amplitude $\Ostray$ couples $\ket a$ to $\ket e$.
Applying a deliberate \emph{holding field} $\Oh$ on the empty $\ket g$--$\ket e$
transition makes the stored state part of a new dark state: the two couplings
form their own $\Lambda$ link to $\ket e$, with dark combination
$\Ostray\ket g-\Oh\ket a$, which aligns with $\ket a$ as $\Oh\gg\Ostray$, so by
the same source identity~\eqref{eq:source} the charge decouples from $\ket e$.
The fast self-discharge is then suppressed as
\begin{equation}
\text{self-discharge rate}\ \propto\ \Big(\frac{\Ostray}{\Oh}\Big)^2 ,
\label{eq:storeprot}
\end{equation}

\begin{figure}[t]
\centering
\includegraphics[width=0.95\columnwidth]{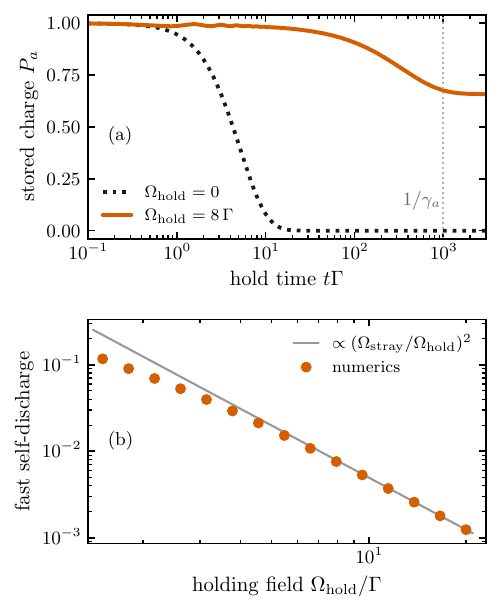}
\caption{\label{fig:store}\textbf{Dark-state protection of stored charge.} After
charging, $\ket a$ is exposed to a parasitic coupling $\Ostray$ on the lossy
$a$--$e$ transition. A holding field $\Oh$ on the empty $g$--$e$ transition
rotates the dark state onto $\ket a$. (a)~Stored charge versus hold time: without
protection ($\Oh=0$) the charge decays radiatively within $\sim\!10/\Gamma$; a
holding field locks it into the dark state, where it persists until the intrinsic
metastable lifetime $1/\gamma_a$. (b)~The residual fast self-discharge falls as
$(\Ostray/\Oh)^2$ (slope-2 line). Parameters $\Ostray=0.5\Gamma$, $\Delta=0$.}
\end{figure}

[Fig.~\ref{fig:store}(b)], converting the radiative decay into the slow intrinsic
metastable rate $\gamma_a$ [Fig.~\ref{fig:store}(a)]. The mechanism is that of a
decoherence-free subspace~\cite{Lidar1998,Liu2019}, but \emph{actively dressed}:
the same dark direction that carries no dissipation during charging shields the
stored charge afterward. There is an optimal holding power, since a stronger
field suppresses the fast channel more while a weaker field re-pumps decayed
population back into the dark state faster (a broader dark resonance); balancing
the two gives $\Oh\sim\Ostray\sqrt{\Gamma/\gamma_a}$ for maximal indefinite retention
(Appendix~\ref{app:storage}). The holding field is dark to the charged state and
performs no work on it; its only cost is the external power to sustain the beam,
an apparatus overhead rather than a drain on the cell. Inhibiting radiative
self-discharge by environment dressing parallels recent non-Markovian
self-discharge control~\cite{XuSelf2024} but follows here from the exact source
identity.

\begin{figure}[t]
\centering
\includegraphics[width=0.95\columnwidth]{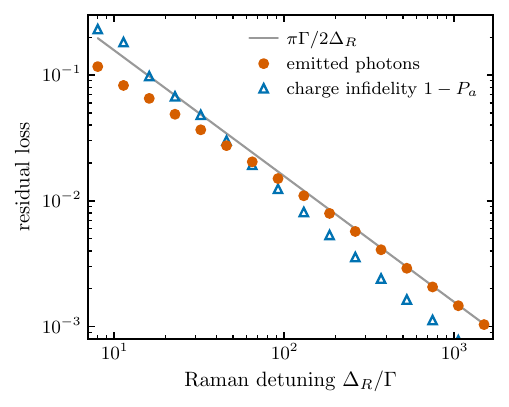}
\caption{\label{fig:compare}\textbf{Cost of a Raman-synthesized counterdiabatic
drive.} When the $\ket g$--$\ket a$ transition is dipole-forbidden and the
counterdiabatic coupling is synthesized as a two-photon Raman process through
$\ket e$ at one-photon detuning $\Delta_R$, the virtual occupation of $\ket e$
scatters. The emitted-photon number and the charge infidelity $1-P_a$ both follow
$\pi\Gamma/2\Delta_R$ (line), vanishing only as $\Delta_R\to\infty$. The exact
$\Psc=0$ holds for an independent, loss-free $\ket g$--$\ket a$ control; a Raman
realization incurs this parametrically small residual. Charging time $T\Gamma=1$.}
\end{figure}
\section{Synthesizing the counterdiabatic drive: the Raman cost}
\label{sec:raman}

The lossless theorem assumes the control $\Omega_a$ is an \emph{independent} field
on the $\ket g$--$\ket a$ transition. Where that transition is directly
addressable---radio-frequency- or microwave-coupled hyperfine or Zeeman
sublevels, or a two-photon-allowed clock transition---this is immediate, and a
direct $\pi$ pulse on $\ket g$--$\ket a$ is itself loss-free; the dark-state route
then earns its keep by charging \emph{fast} when the accessible $\ket g$--$\ket a$
coupling is the counterdiabatic field while strong optical fields set the mixing
angle. Often, however, the $\ket g$--$\ket a$ transition is dipole-forbidden, and
the control must be \emph{synthesized} as a two-photon Raman process through the
very state $\ket e$ we are keeping empty. Driving $\ket g$--$\ket a$ through
$\ket e$ at a large one-photon detuning $\Delta_R$ produces an effective coupling
at the price of a small virtual occupation of $\ket e$, which scatters.
Eliminating $\ket e$ in the Raman leg leaves a residual excited population of
order $|\Omega_a/2\Delta_R|^2$, and inserting this into Eq.~\eqref{eq:psc} over a
charge that sweeps $\theta$ through $\pi/2$ gives an emitted-photon number and a
charge infidelity that both approach
\begin{equation}
1-P_a\ \approx\ \Psc\ \approx\ \frac{\pi\Gamma}{2\Delta_R},
\label{eq:ramanfloor}
\end{equation}
independent of the charging time and vanishing only as $\Delta_R\to\infty$
[Fig.~\ref{fig:compare}]. The exact $\Psc=0$ thus describes the ideal of an
independent, loss-free $\ket g$--$\ket a$ control; a Raman realization replaces it
by this small, quantified residual. The two routes define the central practical
choice: a \emph{direct} control gives provable zero loss for an arbitrary
reservoir but needs a microwave, radio-frequency, or two-photon-clock drive,
while a \emph{synthesized} Raman control is all-optical but pays the floor of
Eq.~\eqref{eq:ramanfloor}.

\section{Experimental realization}
\label{sec:platforms}

The requirements are modest: a long-lived storage state $\ket a$ at energy
$\hbar\omega_a$, an optically excited bridge $\ket e$ of known linewidth $\Gamma$,
controllable pump and Stokes fields to set the mixing angle $\theta(t)$, and---for
the exact lossless limit---an independently controllable $\ket g$--$\ket a$
coupling to serve as $\Omega_a=2\dot\theta$. By Eq.~\eqref{eq:robust} an amplitude
match at the percent level already suppresses dissipation to the $10^{-4}$ level.
Representative parameters are collected in Table~\ref{tab:platforms}.

Neutral strontium and ytterbium carry optical clock transitions whose upper state
is the storage state $\ket a=\,^3\!P_0$, metastable with lifetimes of order
$10$--$10^2$~s~\cite{Ludlow2015}; the stored quantum is optical,
$\hbar\omega_a\!\approx\!1.8$--$2.1$~eV (the $698$~nm clock line of Sr and the
$578$~nm line of Yb), held for seconds against a microsecond
charge. The bridge is an ordinary excited state with $\Gamma/2\pi$ from kHz to
tens of MHz, and the $\ket g$--$\ket a$ control is a two-photon clock drive or a
Raman synthesis. Trapped ions offer metastable $D_{5/2}$ states---$^{40}$Ca$^+$
(729~nm, lifetime $\sim\!1.1$~s) or $^{88}$Sr$^+$ (674~nm, $\sim\!0.4$~s)---with
strong dipole $P$ levels ($\Gamma/2\pi\sim20$~MHz) as the bridge and
single-ion addressing~\cite{Bruzewicz2019}. Superconducting transmons realize the
$\Lambda$ system in their lowest three GHz-scale levels, where STIRAP and its
superadiabatic accelerations are
established~\cite{Kumar2016,fmodSTIRAP2023,Vepsalainen2019,Krantz2019}; the
transmon is the cleanest setting for the exact theorem, since the
$\ket g$--$\ket a$ coupling can be engineered directly, removing the Raman
residual, and charging is fast (tens to hundreds of ns). Defect centers such as
the nitrogen-vacancy center provide long-lived ground-state spin sublevels with
optical bridges and a directly addressable microwave $\ket g$--$\ket a$
control~\cite{Doherty2013}.

\begin{table*}[t]
\centering
\caption{\label{tab:platforms}Representative (order-of-magnitude) parameters for
candidate quantum-battery cells. $\hbar\omega_a$ is the stored quantum,
$1/\gamma_a$ the intrinsic storage time of $\ket a$, $\Gamma/2\pi$ the bridge
linewidth, $T$ a representative charging time, and $P=\hbar\omega_a/T$ the
resulting per-cell charging power. ``Direct'' control means an independently
addressable $\ket g$--$\ket a$ field, for which the lossless limit
[Eq.~\eqref{eq:theorem}] and its reservoir independence [Eq.~\eqref{eq:bathfree}]
are exact; ``Raman'' means a synthesized control, for which the residual of
Eq.~\eqref{eq:ramanfloor} applies.}
\footnotesize
\setlength{\tabcolsep}{4pt}
\begin{ruledtabular}
\begin{tabular}{lccccccc}
Platform & $\ket a$ & $\hbar\omega_a$ & $1/\gamma_a$ & $\Gamma/2\pi$ & $\ket g$--$\ket a$ control & $T$ & $P$/cell\\
\hline
Neutral Sr/Yb & clock $^3\!P_0$ & $1.8$--$2.1$~eV & $10$--$10^2$~s & $10^{-2}$--$30$~MHz & 2-photon/Raman & $0.1$--$10~\mu$s & $\sim\!10^{-13}$~W\\
Ion Ca$^+$/Sr$^+$ & $D_{5/2}$ & $\sim\!1.7$~eV & $0.4$--$1$~s & $\sim\!20$~MHz & quadrupole/Raman & $1$--$100~\mu$s & $\sim\!10^{-14}$~W\\
Transmon & engineered & $10$--$30~\mu$eV & $10$--$10^2~\mu$s & $\sim$~MHz & microwave (direct) & $10$--$500$~ns & $\sim\!10^{-17}$~W\\
NV/defect & spin sublevel & $\sim\!10~\mu$eV & ms--s & $\sim\!10$~MHz & microwave (direct) & $\sim\!\mu$s & $\sim\!10^{-18}$~W\\
\end{tabular}
\end{ruledtabular}
\end{table*}

A single cell stores one quantum $\hbar\omega_a$; an array of $N$ stores
$N\hbar\omega_a$, with the per-cell powers above scaling linearly with $N$.
Collective coupling to a common drive is the natural route to the enhanced
charging power of the collective-battery
literature~\cite{Binder2015,Campaioli2017,Ferraro2018,Andolina2019}, now without
the dissipative penalty that fast collective driving would otherwise incur, since
each cell charges losslessly. The stored energies are microscopic, so the device
is not a bulk reservoir but a fast, loss-free, addressable energy-storage
primitive and a clean testbed for the thermodynamics of charging, in which the
present scheme fixes the dissipation contribution at zero.

\section{Discussion}
\label{sec:discussion}

Three points delimit the scope. First, the exact $\Psc=0$ and its reservoir
independence require an independent, loss-free $\ket g$--$\ket a$ control; a Raman
synthesis trades exactness for the small floor~\eqref{eq:ramanfloor}, and a
finite two-photon detuning $\delta$ degrades the cancellation smoothly through the
uncancellable imaginary quadrature of the residual source. Second, where the
$\ket g$--$\ket a$ transition is directly and coherently driveable, a bare $\pi$
pulse is already loss-free, and indeed in this idealized limit the
counterdiabatic shortcut and a bare $\pi$ pulse \emph{coincide}: both ride a state
with no $\ket e$ component, and the optical fields and the bridge play no
dynamical role. The dark-state construction earns its keep on three counts that a
bare $\pi$-pulse argument does not supply. (i)~It remains exactly lossless when the
transfer is driven by \emph{strong optical fields} that set the mixing angle
$\theta(t)$ while the controlled $\ket g$--$\ket a$ coupling cancels the
source---the regime relevant when the fast charging speed is set optically rather
than by the (often weak) direct $\ket g$--$\ket a$ Rabi frequency. (ii)~Its
suppression of dissipation is quadratically insensitive to amplitude and
two-photon-detuning errors [Eqs.~\eqref{eq:robust},~\eqref{eq:detuning}].
(iii)~The reservoir-independence theorem of Sec.~\ref{sec:reservoir} establishes
exact losslessness for an \emph{arbitrary} bath through the optical bridge---a
statement that has no bare-$\pi$-pulse counterpart, since the optical route is the
one that exposes $\ket e$ to the reservoir in the first place. The substantive
comparison is therefore with the optical alternatives of
Fig.~\ref{fig:charge}, against which the shortcut is the unique protocol that is
simultaneously fast and lossless. Third, the reservoir-independence argument assumes a
vacuum bath coupled through $\ket e$; a finite-temperature bath, or a bath
coupling directly to $\ket a$ or $\ket g$, would reintroduce a channel that the
dark state does not close, and is left for future work. Within these bounds, the
construction is exact and platform-agnostic, and the structural statement---one
bright mode carries all emission, one residual source gates it, and cancelling
that source decouples the system from the reservoir entirely---applies to any
driven open system with a single dominant decay channel.

\section{Conclusion}
\label{sec:conclusion}

We have shown that a quantum battery can be charged quickly and \emph{exactly}
losslessly, and that the losslessness is independent of the reservoir. The
no-jump identity $\Psc=\Gamma\int|c_e|^2dt$ ties dissipation to the population of
the lossy bridge; the source identity $\Omega_P c_g+\Omega_S c_a=\Omega b$
isolates the single bright amplitude that feeds it; and the residual source
$F=\dot\theta-\Omega_a/2$ is the single quantity joining the protected dark sector
to the lossy one. The counterdiabatic control $\Omega_a=2\dot\theta$ at two-photon
resonance makes $F\equiv0$, so the bridge stays rigorously empty, the cell charges
to full ergotropy $\hbar\omega_a$ with no photon emitted at any speed, and---because
the dark state is decoupled from the system--reservoir coupling itself---the
emission is exactly zero for an arbitrary bath spectral density, Markovian or not.
The bright sector carries the entire non-Hermitian structure, including a
Markovian second-order exceptional point that reservoir memory promotes to a
third-order one, and the lossless protocol is by construction exempt from the
resulting dissipation phase diagram. This inverts the logic of
dissipation-engineered and non-Markovian batteries, which harness exceptional
points and memory as resources: here the cell is charged so that the environment
is never engaged. The same dark direction protects the stored charge with a
self-discharge suppressed as $(\Ostray/\Oh)^2$, the residual dissipation from an
imperfect control is quadratic in the calibration error, and a Raman synthesis of
the control costs only $\pi\Gamma/2\Delta_R$. With representative parameters for
neutral atoms, trapped ions, transmons, and defect centers placing the scheme
within reach of current hardware, the organizing idea---identify the one mode that
radiates and the one source that feeds it, and cancel that source---makes
loss-free, fast charging and protected storage exact, transparent, and
reservoir-independent in a single framework.

\section*{Acknowledgments}
This publication was partially supported by the Qatar Research, Development and Innovation (QRDI) Council under the Academic Research Grant ARG01-0603-230468. The findings and views expressed herein are solely the responsibility of the authors.

\appendix

\section{No-jump unravelling and the emitted-photon probability}
\label{app:nojump}

In the quantum-trajectory picture~\cite{PlenioKnight1998,BreuerPetruccione}, the
smooth evolution between photon emissions is generated by the non-Hermitian
Hamiltonian $\Hnh=H-\tfrac{i}{2}\Gamma\ketbra e e$ obtained by discarding the
recycling term of the master equation. For a conditional state $\ket\psi$ evolved
by $\Hnh$ from a normalized initial state, the norm obeys
$\tfrac{d}{dt}\langle\psi|\psi\rangle=-\Gamma|c_e|^2$, since
$\Hnh-\Hnh^\dagger=-i\Gamma\ketbra e e$. The norm lost up to time $T$ is the
probability that a photon was emitted, so
$\Psc=1-\langle\psi(T)|\psi(T)\rangle=\Gamma\int_0^T|c_e|^2dt$, which is
Eq.~\eqref{eq:psc}. The relation is exact and makes no assumption about the
magnitude of $c_e$.

\section{Terminology: dark/bright states and dark/bright basis}
\label{app:darkbasis}

The dark and bright \emph{states} are the instantaneous Hamiltonian-selected
superpositions
\begin{equation}
\ket D=\cos\theta\,\ket g-\sin\theta\,\ket a,\qquad
\ket B=\sin\theta\,\ket g+\cos\theta\,\ket a,
\end{equation}
where $\tan\theta=\Omega_P/\Omega_S$. They are named by their optical coupling:
$H_{\rm opt}\ket D=0$, while $H_{\rm opt}\ket B=(\Omega/2)\ket e$. Thus
$\ket D$ is dark because it does not feed the radiative state, whereas $\ket B$
is bright because it does.

The dark/bright \emph{basis} is instead the representation
$\{\ket D(t),\ket B(t),\ket e\}$ in which the state is written as
$\ket\psi=d\ket D+b\ket B+c_e\ket e$. Since this basis moves with $\theta(t)$,
it generates the geometric coupling
$\ket{\dot D}=-\dot\theta\ket B$ and $\ket{\dot B}=\dot\theta\ket D$. Hence a
state can be instantaneously dark with respect to the optical Hamiltonian while
still leaking into the bright sector because the basis itself is changing. The
counterdiabatic field cancels precisely this basis-motion-induced coupling.

\section{Damped-pseudomode construction}
\label{app:pseudomode}

A system coupled to a reservoir with a Lorentzian spectral density,
\begin{equation}
J(\omega)=\frac{1}{2\pi}\,\frac{\Gamma\lambda^2}{(\omega-\omega_e)^2+\lambda^2},
\end{equation}
of width $\lambda$ and weight fixed by $\Gamma$, is reproduced \emph{exactly} by
coupling the system to a single harmonic pseudomode that itself decays into a flat
(Markovian) bath at rate $\lambda$~\cite{Garraway1997}. The system's reduced
dynamics is identical to that of the original structured reservoir; no Born or
secular approximation is made. In the single-excitation, no-jump sector the
conditional amplitudes obey Eq.~\eqref{eq:pseudo} with system--pseudomode coupling
$g$ related to the reservoir weight by $\Gamma=2g^2/\lambda$. The only
anti-Hermitian term is the pseudomode damping $-i\lambda A$, so the conditional
norm obeys $\tfrac{d}{dt}(|d|^2+|b|^2+|c_e|^2+|A|^2)=-2\lambda|A|^2$, and the
emitted-photon probability is the norm lost from the pseudomode,
$\Psc=2\lambda\int_0^T|A|^2dt$. As $\lambda\to\infty$ at fixed $\Gamma=2g^2/\lambda$
the pseudomode follows the excited amplitude adiabatically,
$A\simeq-ig\,c_e/\lambda$, and $gA\to-i(\Gamma/2)c_e$, recovering the Markovian
$i\dot c_e=(\Omega/2)b-i(\Gamma/2)c_e$ and $\Psc=\Gamma\int|c_e|^2dt$. The memory
time of the reservoir is $\lambda^{-1}$.

\section{Spectral structure: Markovian EP2 and structured-reservoir EP3}
\label{app:ep}

\emph{Markovian core.} Eliminating the pseudomode leaves the damped two-level
generator of $(b,c_e)$,
$M_2=\big(\begin{smallmatrix}0 & \Omega/2\\ \Omega/2 & -\zeta\end{smallmatrix}\big)$
with $\zeta=\Delta+i\Gamma/2$, whose eigenvalues are
$\lambda_\pm=\tfrac12(-\zeta\pm\sqrt{\zeta^2+\Omega^2})$. At $\Delta=0$ the root is
$\sqrt{\Omega^2-\Gamma^2/4}$, so the two complex frequencies coalesce at
$\Omega=\Gamma/2$, where $\lambda_+=\lambda_-=-i\Gamma/4$, $M_2$ is a Jordan block,
and a residual excited amplitude decays as $c_e\propto t\,e^{-\Gamma t/4}$---the
critical damping of a damped oscillator and a second-order exceptional
point~\cite{Heiss2012,MiriAlu2019}. Strong driving ($\Omega>\Gamma/2$)
under-damps this channel; weak driving over-damps it.

\emph{Structured-reservoir core.} With the pseudomode retained, the generator is
$M_3$ of Eq.~\eqref{eq:M3}, with characteristic polynomial
$\mu^3+i\lambda\mu^2-(g^2+\Omega^2/4)\mu-i\lambda\Omega^2/4=0$. A third-order
exceptional point requires a triple root $\mu_0$; matching $(\mu-\mu_0)^3$ to the
polynomial gives $\mu_0=-i\lambda/3$, $g^2+\Omega^2/4=\lambda^2/3$, and
$\Omega^2=4\lambda^2/27$, i.e.
\begin{equation}
\small
\Omega=\frac{2\lambda}{3\sqrt3},\quad g^2=\frac{8\lambda^2}{27},\quad
\text{equivalently}\quad \lambda=\frac{27}{16}\Gamma,\ \ \Omega\approx0.65\,\Gamma,
\end{equation}
at which all three eigenvalues meet at $\mu_0=-i\lambda/3$ and a residual excited
amplitude decays as $t^2\,e^{-\lambda t/3}$. The Markovian EP2 at $\Omega=\Gamma/2$
is the $\lambda\to\infty$ endpoint of the exceptional ridge that threads this EP3,
as mapped in Fig.~\ref{fig:reservoir}(b). The promotion of an EP2 to an EP3 by
reservoir memory matches the non-Markovian exceptional-point phenomenology
reported in Refs.~\cite{NMEP2025,OptoNMEP2026}.

\section{Protected-storage algebra and the optimal holding field}
\label{app:storage}

With a parasitic coupling $\Ostray$ on $\ket a$--$\ket e$ and a holding field $\Oh$
on $\ket g$--$\ket e$, the two form a $\Lambda$ link to $\ket e$ with dark state
$\ket{D'}\propto\Ostray\ket g-\Oh\ket a$ and bright state
$\ket{B'}\propto\Oh\ket g+\Ostray\ket a$. The charged state decomposes as
$\ket a=(\Ostray\ket{B'}-\Oh\ket{D'})/\sqrt{\Oh^2+\Ostray^2}$, so its overlap with
the lossy bright direction is $\Ostray/\sqrt{\Oh^2+\Ostray^2}\to\Ostray/\Oh$ for
$\Oh\gg\Ostray$. The fast radiative channel, of bare rate $\sim\Gamma$, is
therefore weighted by $(\Ostray/\Oh)^2$, giving Eq.~\eqref{eq:storeprot}. The
bright component decays at $\sim\Gamma$ and is continuously re-pumped toward
$\ket{D'}$ at a rate set by the dressed splitting $\sim\sqrt{\Oh^2+\Ostray^2}$;
balancing the residual leakage $\gamma_{\rm fast}\sim\Gamma(\Ostray/\Oh)^2$ against
the intrinsic metastable loss $\gamma_a$ and the re-pumping gives an optimal
holding amplitude $\Oh\sim\Ostray\sqrt{\Gamma/\gamma_a}$, at which indefinite
retention is maximized. The
holding field is dark to $\ket a$ and performs no work on the stored charge.

\end{document}